\documentclass[12pt]{article}

\usepackage{latexsym}

\usepackage{graphicx}

\begin{document}


\title{Magnetized black holes and black rings in the higher dimensional  dilaton gravity}

\author{
     Stoytcho S. Yazadjiev \thanks{E-mail: yazad@phys.uni-sofia.bg}\\
{\footnotesize  Department of Theoretical Physics,
                Faculty of Physics, Sofia University,}\\
{\footnotesize  5 James Bourchier Boulevard, Sofia~1164, Bulgaria }\\
}

\date{}

\maketitle

\begin{abstract}
In this paper we consider magnetized black holes and black rings  in the higher dimensional
dilaton gravity. Our study is based on exact solutions generated by applying
a Harrison transformation to known asymptotically flat black hole and black ring solutions in higher dimensional
spacetimes. The explicit solutions include the magnetized version of the higher dimensional Schwarzschild-Tangherlini black holes, Myers-Perry black holes and five dimensional (dipole) black rings.  The basic physical quantities of the magnetized objects are calculated. We also discuss  some  properties of the solutions and their thermodynamics. The ultrarelativistic limits of the magnetized solutions are briefly discussed and an explicit example is given
for the $D$-dimensional magnetized Schwarzschild-Tangherlini black holes.\\
{PACS: 04.20.Jb ; 04.50.+h}
\end{abstract}


\sloppy

\section{Introduction}

In recent years the higher dimensional gravity is attracting much interest. Apart from the fact
that the higher dimensional gravity is interesting in its own right, the increasing amount of works
devoted to the study of the higher dimensional spacetimes is inspired from the string theory and
the brane-world scenario with large extra dimensions \cite{AHDD}-\cite{KOKO}.  This scenario suggests a  possibility of unification of the
electro-weak and Planck scales at TeV scale. A striking prediction in this scenario is the
formation of higher dimensional black holes smaller than the size of the extra dimensions
at accelerators \cite{BF},\cite{GT}.

Some solutions of the higher dimensional classical general relativity have been known for some time. These include
the higher dimensional analogues of  Schwrazschild and Reissner-Nordstrom solution found by Tangherlini \cite{T}
and the higher dimensional generalization  of the Kerr solution found by Myers and Perry \cite{MP}.
As one should expect and as it was confirmed by recent investigations, the gravity in higher dimensions
exhibits much richer dynamics than in four dimensions.An interesting development in the black holes
studies is the discovery of the black ring solutions
of the five-dimensional Einstein equations by Emparan and Reall \cite{ER1}, \cite{ER2}. These are
asymptotically flat solutions with an event horizon of topology $S^2\times S^1$ rather the
much more familiar $S^3$ topology. Moreover, it was shown in \cite{ER2} that both the black hole
and the the black ring can carry the same conserved charges, the mass and a single angular
momentum, and therefore there is no uniqueness theorem in five dimensions. Since the Emparan and
Reall's discovery many explicit examples of black ring solutions were found in various gravity
theories \cite{E}-\cite{P}. Elvang was able
to apply Hassan-Sen transformation to the solution \cite{ER2} to find a charged black ring in the
bosonic sector of the truncated heterotic string theory\cite{E}. A supersymmetric black ring in
five-dimensional minimal supergravity was derived in \cite{EEMR1} and then generalized to the
case of concentric rings in \cite{GG1} and \cite{GG2}. A static black ring solution of the five
dimensional Einstein-Maxwell gravity was found by Ida and Uchida in \cite{IU}. In \cite{EMP}
Emparan derived "dipole black rings" in Einstein-Maxwell-dilaton (EMd) theory
in five dimensions. In this work Emparan showed that the black rings can exhibit novel
feature with respect to the black holes. Black rings can also carry nonconserved charges
which can be varied continuously without altering the conserved charges. This fact leads to
continuous non-uniqness. The thermodynamics of the dipole black rings, within the quasilocal
counterterm method, was discussed by Astefanesei and Radu \cite{AR}. Static and asymptotically
flat black ring solutions in five-dimensional
EMd gravity with arbitrary dilaton coupling parameter $\alpha$ were presented in \cite{KL}.
A systematical derivation of the asymptotically
flat static black ring solutions in five-dimensional EMd gravity with an arbitrary dilaton coupling parameter was given in \cite{Y}. In the same paper and in \cite{Y1}, the author systematically derived new type static and rotating black ring solutions which are not asymptotically flat.

In the present paper we study higher dimensional black holes and black rings immersed in external magnetic fields within the framework of EMd gravity. The interest in  studying black holes under the influence of
external fields has a long history. In 1976 Ernst \cite{ERN} applied a Harrison transformation \cite{HAR} to
Schwarzschild solution to obtain a static black hole in the Melvin universe \cite{BON},\cite{MEL}. The Ernst-Schwarzschild solution
was subsequently discussed by many authors \cite{WK}-\cite{O1}. The Ernst result was generalized to more complicated
metrics as the Kerr-Newman metrics \cite{EW},\cite{DIAZ}. The investigation of the magnetized Kerr-Newman metrics resulted in finding interesting astrophysical effects, such  as charge accretion and flux expulsion from extreme black holes \cite{EW},\cite{DOK}-\cite{KB}. The flux expulsion was also studied in Kaluza-Klein and string theories \cite{CEG}. Five dimensional
black holes in external electromagnetic fields were discussed by Frolov and Aliev \cite{AF} and by Ida and Uchida in \cite{IU}. In \cite{AF} the authors use  Wald's test field approach \cite{WALD} while the discussion in \cite{IU} is based
on exact solutions. Magnetized static and rotating black holes in arbitrary dimensions as well as magnetized
rotating black rings in five dimensions were recently studied by Ortaggio in \cite{O2} within the $D$-dimesional
Einstein-Maxwell (EM) gravity. The discussion is based on exact solutions found by applying a Harrison transformation to known exact black hole and black ring solutions. In this work
 Ortaggio also discussed some properties of the magnetized black holes and black rings as well as their thermodynamics  and gave the utrarelativistic limit of the magnetized $D$-dimensional Schwarzshild solution. Here we generalize the results of \cite{O2} in the presence of dilaton field non-minimally coupled to the electromagnetic field.

The paper is organized as follows. In the first section we systematically derived the Harrison transformation
for the EMd equations in $D$-dimensional  spacetimes with relevant symmetries. Then in the subsequent sections
we apply the Harrison transformation to known black hole and black ring solutions to obtain their magnetized
versions. We also discuss some properties of the magnetized solutions as well as their thermodynamics.
The last section is devoted to summary of the results. In Appendix B we present the explicit expressions of
the utrarelativistic limits of some of the magnetized solutions.

\section{Basic equations and Harrison transformation}

The EMd gravity in $D$-dimensional spacetimes is described by the action

\begin{equation}\label{EMDA}
S= {1\over 16\pi} \int d^Dx \sqrt{-g}\left(R -
2g^{\mu\nu}\partial_{\mu}\varphi \partial_{\nu}\varphi  -
e^{-2\alpha\varphi}F^{\mu\nu}F_{\mu\nu} \right).
\end{equation}

The field equations derived from this action are

\begin{eqnarray}
R_{\mu\nu} &=& 2\partial_{\mu}\varphi \partial_{\nu}\varphi + 2e^{-2\alpha\varphi} \left[F_{\mu\rho}F_{\nu}^{\rho} - {g_{\mu\nu}\over 2(D-2)} F_{\beta\rho} F^{\beta\rho}\right], \\
\nabla_{\mu}\nabla^{\mu}\varphi &=& -{\alpha\over 2} e^{-2\alpha\varphi} F_{\nu\rho}F^{\nu\rho}, \\
&\nabla_{\mu}&\left[e^{-2\alpha\varphi} F^{\mu\nu} \right]  = 0 .
\end{eqnarray}

We consider spacetimes admitting a spacelike, hypersurface-orthogonal Killing vector which
we denote by $\eta$. In adapted coordinates in which $\eta=\partial/\partial y$, the spacetime
metric can be written in the form

\begin{equation}
ds^2 = e^{2u}dy^2 + e^{-{2u\over D-3}} h_{ij}dx^idx^j
\end{equation}

where $h_{ij}$ is a $(D-1)$-dimensional metric with Lorentz signature.Both $u$ and $h_{ij}$
depend on the coordinates $x^i$ only. The electromagnetic field
is taken in the form

\begin{equation}
F = dA_{y}\wedge dy.
\end{equation}

The potential $A_{y}$ depends on $x^{i}$ only.  In terms of the potentials $u$,$A_{y}$
and $\varphi$ the field equations read

\begin{eqnarray}
{\cal D}_{i}{\cal D}^{i}u =
- 2{D-3\over D-2} e^{-2\alpha\varphi -2u}h^{ij}{\cal D}_{i}A_{y} {\cal D}_{j}A_{y},\\
{\cal D}_{i}{\cal D}^{i}\varphi =
- \alpha e^{-2\alpha\varphi -2u}h^{ij}{\cal D}_{i}A_{y} {\cal D}_{j}A_{y},\\
{\cal D}_{i}\left(e^{-2\alpha\varphi - 2u}{\cal D}^{i}A_{y} \right) = 0, \\
R(h)_{ij}= {D-2\over D-3}\partial_{i}u\partial_{j}u + 2\partial_{i}\varphi\partial_{j}\varphi
+ 2e^{-2\alpha\varphi - 2u}\partial_{i}A_{y}\partial_{j}A_{y}.
\end{eqnarray}

Here ${\cal D}_{i}$ and $R(h)_{ij}$ are the coderivative operator and Ricci tensor with respect
to the metric $h_{ij}$. These equations can be derived from the action

\begin{eqnarray}\label{DRA}
S = \int d^{D-1}x\sqrt{|h|}\left[R(h) - {D-2\over D-3}h^{ij}\partial_{i}u\partial_{j}u
- 2h^{ij}\partial_{i}\varphi\partial_{j}\varphi -
2e^{-2\alpha\varphi - 2u}h^{ij}\partial_{i}A_{y}\partial_{j}A_{y} \right].
\end{eqnarray}

In order to unveil the symmetries of the action (\ref{DRA}) we introduce the symmetric
matrix
\begin{eqnarray}
P = e^{(\alpha_{D} -1)u} e^{- (\alpha_{D} + 1)\varphi_{D}}\left(%
\begin{array}{cc}
  e^{2u + 2\alpha_{D}\varphi_{D}} + (1 + \alpha^2_{D})\Psi^2_{D} & - \sqrt{1 + \alpha^2_{D}}\Psi_{D} \\
- \sqrt{1 + \alpha^2_{D}}\Psi_{D}  &  1 \\\end{array}%
\right)
\end{eqnarray}

where

\begin{equation}
\alpha_{D}= \sqrt{{D-2\over 2(D-3)}} \alpha, \,\,\,
\varphi_{D} = \sqrt{{2(D-3)\over (D-2)}} \varphi, \,\,\,
\Psi_{D}= \sqrt{{2(D-3)\over (D-2)}} A_{y} .
\end{equation}

The the action (\ref{DRA}) can be written in the form of a $\sigma$-model action

\begin{equation}
S =\int d^{D-1}x\sqrt{|h|}\left[R(h)
+ {1\over 2(1+ \alpha_{D}^2) }{(D-2)\over (D-3)} h^{ij}
Sp \left({\cal D}_{i}P {\cal D}_{i}P^{-1} \right)  \right].
\end{equation}

Clearly, the action is invariant under the group $GL(2,R)$ where the natural group action
is

\begin{equation}
P \to GPG^T.
\end{equation}

In fact, the matrix $P$ parameterizes the coset $GL(2,R)/SO(2)$.  A similar $\sigma$-model
was previously discussed in \cite{GR}. However, in \cite{GR}, the target space is parameterized
by $3\times 3$ matrices while here we give the target space parameterization in terms of
$2\times 2$ matrices.

In what follows we will be interested in a particular subgroup of $SL(2,R)\subset GL(2,R)$
which gives the Harrison transformation, namely the subgroup consisting of the matrices

\begin{eqnarray}
H = \left(%
\begin{array}{cc}
  1 & 0 \\
  b & 1 \\\end{array}%
\right) .
\end{eqnarray}

The Harrison transformation generates new solutions from known ones which have the same
$(D-1)$-dimensional metric $h_{ij}$ and the new matrix

\begin{equation}
P^{\prime} = HPH^{T}.
\end{equation}

In explicit form the new potentials are given by

\begin{eqnarray}
e^{2u^{\prime}} &=& \Lambda^{-{2\over 1+\alpha^2_{D} }}e^{2u} ,\\
e^{-2\varphi^{\,\prime}_{D}} &=& \Lambda^{{2\alpha_{D}\over 1+\alpha^2_{D} }}e^{-2\varphi_{D}} ,\\
\Psi^{\,\prime}_{D} &=&\Lambda^{-1} \left[\Psi_{D}
+ {b\over \sqrt{1 + \alpha^2_{D}} }\left(e^{2u +2\alpha_{D}\varphi_{D}}
+ (1+\alpha^2_{D})\Psi^2_{D} \right) \right],
\end{eqnarray}

where

\begin{equation}
\Lambda = b^2 e^{2u + 2\alpha_{D}\varphi_{D}}
+ \left(1 + b\sqrt{1 + \alpha^2_{D}}\Psi_{D} \right)^2.
\end{equation}

In other words, the old metric

\begin{equation}
ds^2 = e^{2u}dy^2 + g_{ij}dx^idx^j
\end{equation}

is transformed to the new one

\begin{equation}
d{s^{\,\prime}}^2 = e^{2u^{\,\prime}}dy^2
+ \Lambda^{{2\over (D-3)(1+ \alpha^2_{D}) }} g_{ij}dx^idx^j.
\end{equation}

In the particular case $\alpha=0$, we obtain the Harrison transformation in the Einstein-Maxwell
gravity  discussed in \cite{O2}. For $D=4$ we recover the Harrison transformation in the four-dimensional
EMd gravity \cite{DGKT}.

\section{Dilaton-Melvin solution}

In this section we derive and briefly comment on the dilaton-Melvin solution which plays the role of
background for all magnetized objects we consider in this work. The dilaton-Melvin solution in $D$ dimensions was first found in \cite{GM} by solving the corresponding
equations. In order to derive this solution here we  apply the Harrison transformation to the $D$-dimensional
flat spacetime presented in appropriate coordinates

\begin{equation}
ds^2 = -dt^2 + dz^2_{1} + dz^2_{2} + ... + dz^2_{D-3} + d\rho^2  + \rho^2 d\phi^2.
\end{equation}

The Harrison transformation with respect to the Killing vector $\partial/\partial \phi$ then generates the
dilaton-Melvin solution

\begin{eqnarray}\label{DMS}
ds^2 &=& \Lambda^{{2\over (D-3)(1+\alpha^2_{D})}}
\left[-dt^2 + dz^2_{1} + dz^2_{2} + ...dz^2_{D-3} + d\rho^2 \right] +
\Lambda^{-{2\over 1+ \alpha^2_{D}}}\rho^2d\phi^2 ,\nonumber \\
e^{-2\alpha\varphi}&=& \Lambda ^{{2\alpha^2_{D}\over 1+ \alpha^2_{D}}},  \\
A_{\phi} &=&  \Lambda^{-1}\sqrt{{D-2\over 2(D-3)}} {b\over \sqrt{1+\alpha_{D}^2}} \rho^2
\nonumber, \\
\Lambda &=& 1  + b^2\rho^2 .\nonumber
\end{eqnarray}

If $\alpha=0$ this solution is the $D$-dimensional EM Melvin solution,
whose properties were discussed in \cite{GW}(and in \cite{KT}-\cite{KT1} for $D=4$ ).
The properties of the $D$-dimensional dilaton-Melvin solution are similar.
The geometry is a warped product of a $(D-2)$-dimensional Minkowski spacetime
and a noncompact 2-dimensional space $M_{2}$ with a metric

\begin{equation}
dl^2 = \Lambda^{{2\over (D-3)(1+\alpha^2_{D})}} d\rho^2 +
\Lambda^{-{2\over 1+ \alpha^2_{D}}}\rho^2d\phi^2.
\end{equation}

The circumference of the circles $\rho=constant$ at first increases and then monotonically
decreases to zero as $\rho\to \infty$.
The dilaton field $\varphi$ is divergent at $\rho\to \infty$, but  the scalar invariants tends to zero for
$\rho\to \infty$, for example

\begin{equation}
R_{\mu\nu\alpha\beta}R^{\mu\nu\alpha\beta}\sim \rho^{-4{ (D-3)(1+\alpha^2_{D}) + 2\over (D-3)(1+ \alpha^2_{D})} },
\end{equation}

therefore the geometry is well behaved there. Moreover, it can be checked that the curvature scalars are
everywhere regular. As an example we present the Kretchmann scalar:

\begin{eqnarray}
R_{\mu\nu\alpha\beta}R^{\mu\nu\alpha\beta} = {16b^4\over (D-3)^2(1+\alpha^2_{D})^2 }
\Lambda^{-4{(D-3)(1+\alpha^2_{D}) +1\over (D-3)(1+ \alpha^2_{D})} } K_{1}  \nonumber\\ +
{16b^4\over (1+\alpha^2_{D})^2 } {D-2\over (D-3)^2 } \Lambda^{-2{ (D-3)(1+\alpha^2_{D}) + 2\over (D-3)(1+ \alpha^2_{D})} } K_{2},
\end{eqnarray}

where

\begin{eqnarray}
K_{1} &=& \left[(3D-8) - (D-2){1-\alpha^2_{D}\over 1+ \alpha^2_{D} } b^2\rho^2 \right]^2 ,\\
K_{2} &=& {2\over (D-3)(1+ \alpha^2_{D})^2 } \left({b^2\rho^2\over \Lambda} \right)^2 +
\left[1 - 2{b^2\rho^2\over \Lambda} \right]^2
+ \left[1 - {2\over 1 + \alpha^2_{D}}{b^2\rho^2\over \Lambda} \right]^2 .
\end{eqnarray}

\section{Dilaton-Schwarzschild-Melvin spacetimes }

Let us consider the $D$-dimensional, spherically symmetric Schwarzschild-Tangherlini black hole
spacetimes given by the metric \cite{T}

\begin{equation}
ds_{D}^2 = - \lambda(r)dt^2 + {dr^2\over \lambda(r)} + r^2d\Omega^2_{D-2}
\end{equation}

where

\begin{equation}
\lambda(r) = 1 - {\mu\over r^{D-3}}
\end{equation}

and $d\Omega_{D-2}^2$ is the line element of the unit $(D-2)$-dimensional sphere. The parameter
$\mu>0$ is related to the black hole mass via the relation\footnote{$\Omega_{(D-2)}$ is the area
of the unit $(D-2)$-dimensional sphere.}

\begin{equation}
M = {\mu(D-2)\over 16\pi}\Omega_{D-2}.
\end{equation}

It is convenient to present the line element  $d\Omega_{D-2}^2$ in the form

\begin{equation}
d\Omega_{D-2}^2 = \cos^2\theta d\Omega^2_{D-4} + d\theta^2 + \sin^2\theta d\phi^2
\end{equation}

or

\begin{equation}
ds_{D}^2 = - \lambda(r)dt^2 + {dr^2\over \lambda(r)}
+ r^2\cos^2\theta d\Omega^2_{D-4} + r^2d\theta^2 + r^2\sin^2\theta d\phi^2.
\end{equation}

The Killing vector $\partial/\partial\phi$ is spacelike and  hypersurface-orthogonal,
and, therefore, we can consider the Harrison transformation associated with it. Since the
Schwarzschild-Tangherlini is characterized with trivial dilaton and electromagnetic
field we find

\begin{equation}
\Lambda  = 1 + b^2r^2\sin^2\theta.
\end{equation}

The new solution generated by the Harrison transformation is

\begin{eqnarray}\label{MSTBH}
d{s^{\,\prime}}^2 &=& \Lambda^{2\over (D-3)(1+\alpha^2_{D})}\left[
- \lambda(r)dt^2 + {dr^2\over \lambda(r)}
+ r^2\cos^2\theta d\Omega^2_{D-4} + r^2d\theta^2 \right] + \Lambda^{-{2\over 1+\alpha_{D}^2
} }r^2\sin^2\theta d\phi^2 , \nonumber  \\
e^{-2\alpha\varphi^{\,\prime}} &=&
\Lambda ^{{2\alpha^2_{D}\over 1+ \alpha^2_{D}} }, \nonumber \\
A^{\prime}_{\phi} &=& \Lambda^{-1}\sqrt{{D-2\over 2(D-3)}} {b\over \sqrt{1+\alpha_{D}^2} }r^2\sin^2\theta .
\end{eqnarray}

This solution reduces to that of the EM gravity  \cite{O2} for $\alpha=0$.

The constant $b$ introduced  by  the Harrison transformation  parameterizes the strength of the magnetic field.
In order to find the relation between the parameter $b$ and the asymptotic magnetic field $B$ let us consider
the invariant

\begin{equation}
{1\over 2} F_{\mu\nu}F^{\mu\nu} = 2{(D-2)\over (D-3)} {b^2\over  1+ \alpha^2_{D} }
\left[\lambda(r) \sin^2\theta + \cos^2\theta\right] \Lambda^{{2(D-4)\over (D-3)(1+ \alpha^2_{D})}-4} .
\end{equation}

This invariant takes the constant value $B^2$ at the "axis" $\theta=0$

\begin{equation}
B^2 = 2{(D-2)\over (D-3)} {b^2\over  1+ \alpha^2_{D} }
\end{equation}

which gives the relation between the asymptotic magnetic field and the parameter $b$.

The magnetized solution (\ref{MSTBH}) has a single horizon located where $\lambda(r)=0$
i.e. $r_{h}=\mu$. As in the case of EM gtavity the location of the horizon
is not effected by the value of the dilaton and magnetic field.
The horizon is regular and the spacetime can be extended across the
horizon by the standard techniques. The curvature invariants diverge at $r=0$
indicating the presence of a curvature singularity there. As an illustrative example
we may consider the Ricci scalar curvature

\begin{equation}
R ={4b^2\over 1+\alpha^2_{D}}\Lambda^{-2{1 + (D-3)(1+\alpha^2_{D})\over (D-3)(1+\alpha^2_{D})}}
\left[{D-2\over D-3}{\alpha^2_{D}\over 1+\alpha^2_{D}}r^2\sin^2\theta  + {D-4\over D-3}\right]
\left[1 - {\mu\over r^{D-3}}\sin^2\theta \right].
\end{equation}

For $r\to \infty $ the solution tends to the $D$-dimensional  dilaton-Melvin solution (\ref{DMS}), which can be obtained
by setting  $\mu=0$. The connection between the coordinates of (\ref{DMS}) and those of (\ref{MSTBH})
is given by
\begin{eqnarray}
\rho = r\sin\theta ,\,\,\,\\
\left(z^2_{1} + z^2_{2} + ...+ z^2_{D-3} \right)^{1/2}=r\cos\theta ,\\
d(r\cos\theta)^2 + (r\cos\theta)^2 d\Omega^2_{D-4}= dz^2_{1} + dz^2_{2}+ ...+dz^2_{D-3} .
\end{eqnarray}

The solution (\ref{MSTBH}) can be interpreted as a black hole in dilaton-Melvin
background (or dilaton-Melvin Universe). As should be expected, the background
deforms the black hole horizon and its geometry is now different from that of the
round $(D-2)$-dimensional sphere of radius $r_{h}$. The geometry of the horizon  is
given by the line element

\begin{equation}
ds^2_{h} = \Lambda_{h}^{{2\over (D-3)(1+\alpha^2_{D})}} \left[r_{h}^2\cos^2\theta d\Omega^2_{D-4}
+ r_{h}^2d\theta^2 \right] + \Lambda_{h}^{-{2\over 1+ \alpha^2_{D}}}r_{h}^2\sin^2\theta d\phi^2
\end{equation}

where

\begin{equation}
\Lambda_{h}= 1 + b^2r^2_{h}\sin^2\theta .
\end{equation}

A good illustrative measure of the departure form sphericity is the Ricci scalar curvature
of the horizon

\begin{eqnarray}
R_{h} = {\Lambda_{h}^{-2\over (D-3) (1+ \alpha^2_{D})}\over r^2_{h}} \left[ (D-2)(D-3) +
{4b^2r^2_{h}\over 1+ \alpha^2_{D}} {\Lambda_{h}^{-1} \over (D-3) }\left[D\cos^2\theta - (D-3)\sin^2\theta \right]
\right. \nonumber \\ \left. -
{4b^4r^4_{h}\over 1+\alpha^2_{D}} {\Lambda_{h}^{-2}\over (D-3)}
\left[2 + {1\over 1+ \alpha^2_{D}} {D^2 - 5D + 8\over D-3 } \right]\sin^2\theta \cos^2\theta  \right].
\end{eqnarray}

As it can be seen the Ricci scalar curvature differs from that of the
round $(D-2)$-dimensional sphere of radius $r_{h}$, $R_{D-2}=(D-2)(D-3)/r^2_{h}$ .
The background deforms the horizon but preserves the horizon area
since the Harrison transformation leaves the determinant of the horizon metric invariant. Therefore
the horizon area of the magnetized black hole is that of the Schwarzschild-Tangherlini black hole

\begin{equation}
{\cal A}_{h} = \Omega_{D-2}r_{h}^{D-2} .
\end{equation}

The horizon temperature can be found by Eucleadizing the metric and the result is

\begin{equation}
T = {1\over 2\pi } {(D-3)\mu\over  r^{D-2}_{h}} = {1\over 4\pi } {D-3\over r_{h} }
\end{equation}

and is the same as for the Schwarzschild-Tangherlini black hole.

It is interesting to compute the magnetic flux across a portion $\Sigma$ of the horizon. The flux is found to
be

\begin{equation}
\Phi= \oint_{\partial\Sigma}A^{\prime}_{\phi}  = {B\pi r^2_{h}\sin^2\theta\over  1 + {(D-3)\over 2(D-2)} (1+ \alpha^2_{D})B^2r^2_{h}\sin^2\theta  }.
\end{equation}

For weak magnetic fields the flux  is proportional to the magnetic field strength, while it tends to zero
for $B\to \infty$. There is a maximum of the magnetic flux for intermediate values of $B$ which is a consequence of the concentration of the  magnetic field under  its self-gravity.

In order
to compute the black hole mass we  use  the quasilocal formalism
approach (see Appendix A). Following this approach we decompose the metric into the form

\begin{equation}
d{s^{\prime}}^2 = - N^2 dt^2 + \chi_{ij}(dx^i + N^idt)(dx^j + N^jdt)
\end{equation}

where

\begin{eqnarray}
N^2 &=& \Lambda^{2\over (D-3)(1+ \alpha^2_{D}) } \lambda(r) ,\\
N^i &=& 0 ,\\
\chi_{ij}dx^idx^j &=& \Lambda^{2\over (D-3)(1+\alpha^2_{D})}\left[
{dr^2\over \lambda(r)}
+ r^2\cos^2\theta d\Omega^2_{D-4} + r^2d\theta^2 \right]  ,\\
&+& \Lambda^{-{2\over 1+\alpha_{D}^2} }r^2\sin^2\theta d\phi^2 .
\end{eqnarray}

Further we consider the $(D-2)$-dimensional surfaces $S^{r}_{t}$ with the unit spacelike
normal $n_{r}= \sqrt{\chi_{rr}}{\partial\over\partial r}$ and metric

\begin{equation}
\sigma_{ab}dx^ax^b = \Lambda^{2\over (D-3)(1+\alpha^2_{D})}\left[
r^2\cos^2\theta d\Omega^2_{D-4} + r^2d\theta^2 \right]  \\
+ \Lambda^{-{2\over 1+\alpha_{D}^2} }r^2\sin^2\theta d\phi^2 .
\end{equation}

The extrinsic curvature is

\begin{equation}
k = - {(D-2) \sqrt{\lambda(r)}\over r}  \Lambda^{-{1\over (D-3)(1+ \alpha^2_{D})}  } =
\Lambda^{-{1\over (D-3)(1+ \alpha^2_{D})}  } {\tilde k}
\end{equation}

where ${\tilde k}$ is the extrinsic curvature for the Schwarzschild-Tangherlini solution.

The natural background is obviously the dilaton-Melvin spacetime for which

\begin{equation}
k_{0} = - {(D-2)\over r}  \Lambda^{-{1\over (D-3)(1+ \alpha^2_{D})}  } =
 \Lambda^{-{1\over (D-3)(1+ \alpha^2_{D})}  } {\tilde k}_{0}.
\end{equation}

Taking into account that

\begin{equation}
\sqrt{\sigma} N = \sqrt{\lambda(r)} r^{D-2}  \Lambda^{{1\over (D-3)(1+ \alpha^2_{D})}  }\sqrt{\det\Omega_{D-2}}=
\Lambda^{{1\over (D-3)(1+ \alpha^2_{D})}  } \sqrt{{\tilde \sigma}} {\tilde N}
\end{equation}

we find

\begin{equation}
E(r) = {(D-2)\Omega_{D-2}\over 8\pi} r^{D-3}\sqrt{\lambda(r)} \left[1 - \sqrt{\lambda(r)} \right].
\end{equation}

These explicit calculations show that the contribution of the background cancels out
and as a final result we obtain

\begin{equation}
M =\lim_{r\to \infty}E(r)={\mu(D-2)\Omega_{D-2}\over 16\pi}
\end{equation}

which is exactly the Schwarzschild-Tangherlini black hole mass. In this way we see
that the mass of the black hole is not affected by the background. This a general result
for the mass and angular momentum of black holes in dilaton-Melvin background and it is proven in
Appendix A.
Moreover, we have shown that  all  thermodynamical quantities of the black hole  remain the same
independently of the external magnetic field. Moreover, as it shown in Appendix A the physical Euclidian action
of the Schwarzschild-Tangherlini solution and its magnetized version coincide

\begin{equation}
I_{P} = {\tilde I}_{P}.
\end{equation}

Therefore the black hole thermodynamics is not affected by the background just as in the four dimensional case \cite{RADU}.

\section{Dilaton-Myers-Perry-Melvin spacetimes }

The Myers-Perry black holes \cite{MP} are generalizations of the four-dimensional Kerr solution
to higher dimensions. These solutions are described in different forms depending on that  whether
the spacetime is even or odd dimensional. Here we consider the odd dimensional case.
The even dimensional case is treated in the same way. In spacetime with an odd number of
dimensions the Myers-Perry solution with one of spin parameters set to zero (say $a_{1}=0$)
is given by

\begin{eqnarray}
ds^2 = &-&dt^2 + \sum_{i=2}^{(D-1)/2}(r^2 + a^2_{i})(d\mu^2_{i} + \mu^2_{i}d\phi^2_{i})
+ {\mu r^2 \over \Pi F}\left(dt + \sum_{i=2}^{(D-1)/2}a_{i}\mu^2_{i}d\phi_{i} \right)^2
\nonumber \\
&+& { \Pi F \over \Pi - \mu r^2 }dr^2 + r^2 d\mu^2_{1} + r^2\mu^2_{1} d\phi^2_{1}
\end{eqnarray}

where

\begin{eqnarray}
F &=& 1 - \sum_{i=2}^{(D-1)/2} {a^2_{i}\mu^2_{i}\over a^2_{i} + \mu^2_{i} } ,\\
\Pi &=& r^2 \prod_{i=2}^{(D-1)/2} (r^2 + a^2_{i}), \\
\sum_{i=1}^{(D-1)/2}&\mu^2_{i}& = 1.
\end{eqnarray}

The Myers-Perry solution (in odd dimensions) admits $(D-1)/2$ commuting spacial Killing vectors
${\partial/\partial\phi_{i}}$. What is important for us is that the Killing vector
${\partial/\partial\phi_{1}}$ is hypersurface-orthogonal since we have set $a_{1}=0$. Therefore,
we can apply the Harrison transformation associated with  this Killing vector  to the
Myers-Perry solution. Doing so we find

\begin{eqnarray}
 d{s^{\,\prime}}^2 &=&  \Lambda^{2\over (D-3)(1+\alpha^2_{D})}
\left[ -dt^2 + \sum_{i=2}^{(D-1)/2}(r^2 + a^2_{i})(d\mu^2_{i} + \mu^2_{i}d\phi^2_{i})
\right. \nonumber \\
&& \left. + {\mu r^2 \over \Pi F}\left(dt + \sum_{i=2}^{(D-1)/2}a_{i}\mu^2_{i}d\phi_{i} \right)^2
+ { \Pi F \over \Pi - \mu r^2 }dr^2 + r^2 d\mu^2_{1}\right] +
\Lambda^{-{2\over 1+\alpha^2_{D}}} r^2\mu^2_{1}d\phi^2_{1} , \nonumber \\
A^{\prime}_{\phi_{1}} &=&
\Lambda^{-1}\left( {D-2\over 2(D-3)}\right)^{1/2} {b\over \sqrt{1+ \alpha^2_{D}} } r^2\mu^2_{1} ,
  \\
e^{-2\alpha\varphi^{\,\prime}} &=&
\Lambda^{{2\alpha^2_{D}\over 1 + \alpha^2_{D}} }, \nonumber
\end{eqnarray}

where

\begin{eqnarray}
\Lambda = 1 + b^2r^2\mu^2_{1} \nonumber .
\end{eqnarray}

The outer event horizon is determined as the largest (real) root of $g^{rr} = 0$.
In explicit form the equation for the horizon is given by

\begin{equation}
\Pi - \mu r^2= r^2 \left[\prod_{i=2}^{(D-1)/2} (r^2 + a^2_{i}) - \mu  \right] =0.
\end{equation}
and coinsides with that for the Myers-Perry solution. As it is known this equation
has a positive root independent of the magnitude of $a_{i}$ for\footnote{In general,
for even and odd dimensions, we have $D\ge 6$.} $D\ge 7$, i.e. there exist black holes
with arbitrary angular momentum for $D\ge 7$. In five dimensions, however, there is an upper
bound for the angular momentum.
In the limit $r\to \infty$ we obtain  the dilaton-Melvin background
with asymptotic magnetic field

\begin{equation}
B = \left[2{(D-2)\over (D-3)}\right]^{1/2} {b\over \sqrt{1+ \alpha^2_{D}} } .
\end{equation}

Therefore,
the solution can be interpreted as a rotating black hole in dilaton-Melvin background.

The geometry of the horizon is described by the line element

\begin{eqnarray}
ds_{h}^2 =  \Lambda_{h}^{2\over (D-3)(1+\alpha^2_{D})}
\left[\sum_{i=2}^{(D-1)/2}(r_{h}^2 + a^2_{i})(d\mu^2_{i} + \mu^2_{i}d\phi^2_{i})
\right. \nonumber \\
\left.+ {\mu r_{h}^2 \over \Pi_{h} F_{h}}\left(\sum_{i=2}^{(D-1)/2}a_{i}\mu^2_{i}d\phi_{i} \right)^2
 + r_{h}^2 d\mu^2_{1}\right] +
\Lambda_{h}^{-{2\over 1+\alpha^2_{D}}} r_{h}^2\mu^2_{1}d\phi^2_{1}
\end{eqnarray}

where the subscript $h$ means that the corresponding quantity is evaluated on the horizon.
As in the static case, this metric  describes distorted $(D-2)$-dimensional sphere.
The area of the horizon is

\begin{equation}
{\cal A}_{h} = \Omega_{D-2} r_{h} \prod_{i=2}^{(D-1)/2} (r^2_{h} + a^2_{i})
\end{equation}

which coincides with that of the Myers-Perry solution. This is, as we have already mentioned,
a consequence of the fact that the Harrison transformation preserves the volume element of the
horizon metric.

In order to compute mass and the angular momenta of the black hole we use  the quasilocal formalism
given in Appendix A. The contribution of the background
cancels out and the mass, angular momenta  and  horizon angular velocity are the same as for the
Myers-Perry solution :

\begin{eqnarray}
M &=& {\mu(D-2)\over 16\pi} \Omega_{D-2} , \\
J_{i} &=& {\mu a_{i}\over 8\pi}  \Omega_{D-2}, \\
\omega_{i} &=& {a_{i}\over r^2_{h} + a^2_{i} }.
\end{eqnarray}

We also find that the temperature is not affected by the background and is given by

\begin{equation}
T = {1\over 2\pi} \sum_{i=2}^{(D-1)/2} {r_{h}\over r^2_{h} + a^2_{i} }.
\end{equation}

The same is true for the Euclidean action, $I_{P}={\tilde I}_{P}$ (see Appendix A).
As in the static case, although the background deforms the horizon it does  not affect the
black hole thermodynamics. This seems to be a consequence of the fact that the vector potential
is parallel to the non-rotating Killing vector and the magnetic field and the rotation do not couple.  In
more general case when the Harrison transformation is associated with a rotating Killing vector
one should expect that  the external magnetic field will influence the black hole thermodynamics
via the coupling with the rotation, as in four dimensions \cite{RADU}.

\section{Black rings in dilaton-Melvin background}

The black ring metric is given by \cite{ER2}

\begin{eqnarray}\label{RBRS}
ds^2_{5} = -{F(y)\over F(x)} \left(dt + C(\nu,\lambda){\cal R}{1+y\over F(y)}d\psi \right)^2 \nonumber\\
+ {{\cal R}^2\over (x-y)^2 }F(x)\left[-{G(y)\over F(y)}d\psi^2 - {dy^2\over G(y)} + {dx^2\over G(x)}
+ {G(x)\over F(x)}d\phi^2 \right]
\end{eqnarray}

where

\begin{equation}
F(x) = 1 + \lambda x ,\,\,\, G(x) = (1-x^2)(1+\nu x),
\end{equation}

and

\begin{equation}
C(\nu,\lambda)= \sqrt{\lambda(\lambda-\nu){1 +\lambda \over 1-\lambda}}.
\end{equation}

The coordinates $x$ and $y$ vary within the ranges

\begin{equation}
 -1\le x \le 1, \,\,\, -\infty < y \le -1,
\end{equation}

and the parameters $\lambda$ and $\nu$ within

\begin{equation}
0<\nu\le \lambda<1.
\end{equation}

In order to avoid conical singularities at $y=-1$ and $x=-1$ the angular variables must be
identified with periodicity

\begin{equation}\label{RRSP}
\Delta \psi = \Delta \phi = 2\pi {\sqrt{1-\lambda}\over 1 -\nu}.
\end{equation}

To avoid  a conical singularity at $x=1$ the parameters $\lambda$ and $\nu$ must be related as

\begin{equation}\label{RRSPC}
\lambda = {2\nu\over 1 +\nu^2}.
\end{equation}

With these choices, the solution has a regular horizon of topology $S^2\times S^1$ at $y=-1/\nu$
and ergosurface of the same topology at $y=-1/\lambda$. Asymptotic spacial infinity is
at $x\to y\to -1$. The static solution is obtained for $\lambda=\nu$ instead of (\ref{RRSPC}).
The black ring metric admits three Killing vectors
$\partial/\partial t$, $\partial/\partial \psi$ and $\partial/\partial \phi$. The Killing vector
$\partial/\partial \phi$ is spacelike and hypersurface-orthogonal and the associated with it
Harrison transformation gives the following EMd solution

\begin{eqnarray}
d{s^{\prime}}^2 = \Lambda^{{1\over (1+\alpha^2_{5})}}
 \left[- {F(y)\over F(x)}\left(dt + C(\nu,\lambda){\cal R}{1 + y \over F(y)}d\psi  \right)^2
 \nonumber \right.\\ \left.
+ {{\cal R}^2\over (x-y)^2}F(x) \left(- {G(y)\over F(y)}d\psi^2 - {dy^2\over G(y) } +
{dx^2\over G(x) } \right)\right]
+ \Lambda^{-{2\over (1+\alpha^2_{5})}} {\cal R }^2 {G(x)\over (x-y)^2}d\phi^2 ,\\
A^{\prime}_{\phi} = {\sqrt{3}\over 2 }
{\Lambda^{-1} b\over \sqrt{1 + \alpha^2_{5}} } {\cal R}^2 {G(x)\over (x-y)^2 },\\
e^{-2\alpha\varphi^\prime} = \Lambda^{{2\alpha^2_{5}\over 1 + \alpha^2_{5} }} ,
\end{eqnarray}

with

\begin{equation}
\Lambda  = 1 + b^2 {\cal R}^2 {G(x)\over (x-y)^2} .
\end{equation}

The metric has a horizon at $y=-1/\nu$ with topology $S^2\times S^1$ and a ergosurface at
$y=-1/\lambda$ with same topology. The external fields do not affect the location of the horizon
and the ergosurface. Although  the external fields deform the horizon and the ergosurface,
their  topology remains the same  while the geometry is distorted. For $x,y \to -1$ the solution tends to the
five-dimensional dilaton-Melvin background with magnetic field
\begin{equation}
B = {1-\nu\over \sqrt{1-\lambda} }{\sqrt{3}\over 2} {b\over \sqrt{1+ \alpha^2_{D}} }  .
\end{equation}

This can be seen by performing the coordinate transformation

\begin{eqnarray}\label{BRCT}
r\cos\theta &=& \sqrt{{1-\lambda\over 1 -\nu}} {\cal R} {\sqrt{y^2-1}\over x-y}, \\
r\sin\theta &=& \sqrt{{1-\lambda\over 1 -\nu}}{\cal R} {\sqrt{1-x^2}\over x-y}, \\
{\tilde \psi} &=& {1-\nu\over \sqrt{1-\lambda}}\psi, \\
{\tilde \phi} &=& {1-\nu\over \sqrt{1-\lambda}}\phi .
\end{eqnarray}

The solution then can be
interpreted  as a rotating black ring in dilaton-Melvin background.

As in  the previous cases, the background does not affect the black ring
thermodynamics and the physical quantities characterizing the  magnetized black ring are the same as for
the neutral black ring solution:

\begin{eqnarray}
M &=& {3\pi\over 4}{\cal R}^2 {\lambda\over 1 -\nu}, \\
J_{\psi} &=& {\pi {\cal R}^3 \over 2 } {\sqrt{\lambda(\lambda - \nu)(1+\lambda)}\over (1-\nu)^2 },\\
{\cal A}_{h} &=& 8\pi^2 {\cal R}^3 {\sqrt{\nu^3\lambda (1-\lambda^2) }\over (1-\nu^2)(1+\nu)},\\
\omega_{\psi} &=& {1\over {\cal R}}\sqrt{\lambda - \nu\over \lambda(1+\lambda) },\\
T &=& {1+ \nu\over 4\pi {\cal R} } \sqrt{{1-\lambda\over \nu\lambda (1+ \lambda)}}.
\end{eqnarray}

Here $\omega_{\psi}$ is the angular velocity  of the horizon.

\section{Dipole black rings in dilaton-Melvin background }

The dipole black rings are solutions of the EMd gravity equations given by \cite{EMP}

\begin{eqnarray}
ds^2_{5} = -{F(y)\over F(x)} \left({H(x)\over H(y)} \right)^{1\over 1+\alpha^2_{5}} \left(dt + C(\nu,\lambda){\cal R}{1+y\over F(y)}d\psi \right)^2 \nonumber\\
+ {{\cal R}^2 F(x)\over (x-y)^2 }\left(H(x)H^2(y) \right)^{1\over 1+\alpha^2_{5}}
\left[-{G(y)\over F(y)H^{3\over 1+\alpha^2_{5}}(y) }d\psi^2 \right. \nonumber \\
\left. - {dy^2\over G(y)} + {dx^2\over G(x)}
+ {G(x)\over F(x)H^{3\over 1+\alpha^2_{5}}(x)}d\phi^2 \right],\\
e^{-2\alpha\varphi}  = \left({H(x)\over H(y)}\right)^{{2\alpha^2_{5} \over  1+ \alpha^2_{5}}} ,\\
A_{\phi} = {\sqrt{3}\over 2\sqrt{1+\alpha^2_{5}}} C(\nu,-\mu) {\cal R} {1+x\over H(x)} .
\end{eqnarray}

The functions $F(x)$, $G(x)$ and $C(\lambda,\nu)$, the range of the coordinates $x$, $y$ and the parameters $\lambda$, $\nu$ are the same as in case of the neutral black ring.  The function $H(x)$ is given by

\begin{equation}
H(x) = 1 - \mu x
\end{equation}

where $0 \le\mu <1$.

The possible conical singularities at $x=-1$ and $y=-1$ are avoided by setting

\begin{equation}\label{DBRCSC1}
\Delta \psi = \Delta \phi = 2\pi {(1+ \mu)^{3\over 2(1+ \alpha^2_{5}) }\over 1-\nu } \sqrt{1-\lambda}.
\end{equation}

The avoidance of the conical singularity at $x=1$ simultaneously with (\ref{DBRCSC1}) is achieved only if

\begin{equation}
{1-\lambda\over 1+ \lambda } \left({1+\mu \over 1-\mu }\right)^{3\over 1+ \alpha^2_{5} } = \left({1 - \nu\over 1+ \nu} \right)^2 .
\end{equation}

The solution has a regular outer horizon of topology $S^2\times S^1$ at $y=-1/\nu$. There is also an inner horizon
at $y=-\infty$. The metric  can be continued beyond this horizon to positive values of $y$ until $y=1/\mu$ which
is a curvature singularity. The extremal limit when the two horizon coincide is achieved for $\nu=0$.
In addition there is an ergosurface with ring topology at $y=-1/\lambda$.

The dipole black rings carry local magnetic charge \cite{EMP} given by

\begin{equation}
{\cal Q} = {\sqrt{3}{\cal R} \over 2\sqrt{1+ \alpha^2_{5}} } {(1+\mu)^{2-\alpha^2_{5}\over 2(1+ \alpha^2_{5})} \over
(1-\nu)  } \sqrt{{\mu(\mu+\nu)(1-\lambda)\over 1-\mu }} .
\end{equation}

Therefore the dipole rings are specified by the three physical quantities $M$, $J$ and ${\cal Q}$.
The local charge is independent of the mass $M$ and the angular momentum $J$ and is classically continuous parameter.
This implies infinite classical non-uniqueness in five dimensions.

The Harrison transformation generates the following new EMd solution

\begin{eqnarray}
{ds_{5}^{\,\prime}}^2= \Lambda^{1\over 1+ \alpha^2_{5}} \left[-{F(y)\over F(x)} \left({H(x)\over H(y)} \right)^{1\over 1+\alpha^2_{5}} \left(dt + C(\nu,\lambda){\cal R}{1+y\over F(y)}d\psi \right)^2 \right. \nonumber \\
\left. + {{\cal R}^2 F(x)\over (x-y)^2 }\left(H(x)H^2(y) \right)^{1\over 1+\alpha^2_{5}}
\left(-{G(y)\over F(y)H^{3\over 1+\alpha^2_{5}}(y) }d\psi^2  - {dy^2\over G(y)} + {dx^2\over G(x)}
 \right)\right]  \\
+ \Lambda^{-{2\over 1+ \alpha^2_{5} }} {{\cal R}^2G(x)\over (x-y)^2 } \left({H(y)\over H(x) }\right)^{2\over 1+ \alpha^2_{5}} d\phi^2 , \nonumber
\end{eqnarray}

\begin{eqnarray}
A^{\prime}_{\phi} = \Lambda^{-1}\left[A_{\phi} + {\sqrt{3} b\over 2\sqrt{1+
\alpha^2_{5}} } \left[{{\cal R}^2 G(x)\over (x-y)^2 }\left({H(y)\over H(x) }\right)^{2}   +
{4\over 3}(1+ \alpha^2_{5}) A^2_{\phi} \right]   \right] ,
\end{eqnarray}

\begin{eqnarray}
e^{-2\alpha\varphi^{\prime}} = \left({H(x)\over H(y) } \right)^{2\alpha^2_{5}\over 1+ \alpha^2_{5} }
\Lambda^{2\alpha^2_{5}\over 1+ \alpha^2_{5} } ,
\end{eqnarray}

where

\begin{equation}
\Lambda = b^2 {{\cal R}^2 G(x)\over (x-y)^2 }\left({H(y)\over H(x) } \right)^2
+ \left(1 + {2b\over \sqrt{3} }\sqrt{1+\alpha^2_{5}} A_{\phi} \right)^2 .
\end{equation}

To avoid conical singularities at $x=-1$ and $y=-1$ the angular coordinates must have periodicity given by (\ref{DBRCSC1}). The balance between the forces in the ring will be achieved when, in addition, there are no
conical singularities at $x=1$. Since the ring is carrying a local magnetic charge there will be an additional
force caused by the coupling between the local charge and the external magnetic field. This force manifests
itself  by the presence of the external magnetic field strength (via the parameter $b$)
in the equilibrium condition

\begin{equation}\label{DBRCSC2}
{1-\lambda\over 1+ \lambda } \left({1+\mu \over 1-\mu }\right)^{3\over 1+ \alpha^2_{5} }
\Lambda^{-3\over 1+ \alpha^2_{5}}|_{x=1} = \left({1 - \nu\over 1+ \nu} \right)^2 .
\end{equation}

After performing the coordinate transformation (\ref{BRCT}) one can show that for $x\to y\to -1$ the solution
asymptotes the five-dimensional dilaton-Melvin solution with asymptotic magnetic field

\begin{equation}
B =(1+\mu )^{-3\over 2(1+ \alpha^2_{5}) } {(1-\nu)\over \sqrt{1-\lambda} }{\sqrt{3}\over 2} {b\over \sqrt{1+ \alpha^2_{D}} }  .
\end{equation}

The magnetized dipole ring solution has a regular outer horizon of topology $S^2\times S^1$ at $y=-1/\nu$ and
an ergosurface at $y=-1/\lambda$ with the same topology. There is also an inner horizon
at $y=-\infty$. The metric  can be continued beyond this horizon to positive values of $y$ until $y=1/\mu$ which
is a curvature singularity. The extremal limit when the two horizon coincide is achieved for $\nu=0$.

The external magnetic filed does not affect the values of the mass, angular momentum  and the horizon area and they are the same as for the seed solution (see Appendix A)

\begin{eqnarray}
M &=& {3\pi {\cal R}^2\over 4} {(1+ \mu)^{3\over 1+ \alpha^2_{5}}\over 1-\nu } \left(\lambda
+ {1\over 1 + \alpha^2_{5} } {\mu(1-\lambda) \over 1+ \mu} \right), \\
J_{\psi} &=& {\pi {\cal R}^3\over 2 } {(1+ \mu)^{9\over 2(1+ \alpha^2_{5})} \over (1-\nu)^2}
\sqrt{\lambda(\lambda - \nu)(1+ \lambda)},\\
{\cal A}_{h} &=& 8\pi^2 {\cal R}^3 {(1+ \mu)^{3\over 1+ \alpha^2_{5} } \over (1-\nu)^2(1+ \nu) }
\nu^{3\alpha^2_{5} \over 2(1+ \alpha^2_{5}) } (\mu + \nu)^{3 \over 2(1+ \alpha^2_{5})} \sqrt{\lambda(1-\lambda^2)}.
\end{eqnarray}

The same is true also for the angular velocity of the horizon and the temperature

\begin{eqnarray}
\omega_{\psi} = {1\over {\cal R}} (1 + \mu)^{-{3 \over 2(1+ \alpha^2_{5})}} \sqrt{\lambda -\nu\over \lambda (1+ \lambda) } ,\\
T = {1\over 4\pi {\cal R}} {\nu^{{2-\alpha^2_{5} \over 2(1+ \alpha^2_{5})}} (1+ \nu)\over (\mu+ \nu)^{3 \over 2(1+ \alpha^2_{5})} } \sqrt{1 - \lambda \over \lambda (1+ \lambda) }.
\end{eqnarray}

The local charge of the magnetized ring is

\begin{equation}
{\cal Q} = {\sqrt{3}{\cal R} \over 2\sqrt{1+ \alpha^2_{5}} } {(1+\mu)^{2-\alpha^2_{5}\over 2(1+ \alpha^2_{5})} \over
(1-\nu)  } \sqrt{{\mu(\mu+\nu)(1-\lambda)\over 1-\mu }}\Lambda^{-{1\over 2(1+ \alpha^2_{5})}}|_{x=1}.
\end{equation}

The Euclidean action of the magnetized solution is independent of the magnetic field and coincides with
that of the seed solution, $I_{P}={\tilde I}_{P}$( note, however,  that the balance condition
(\ref{DBRCSC2}) does involve the strength of the external magnetic field). At first sight, it seems strange that the
thermodynamics is not affected by the external magnetic field. One might expect that the coupling between
the local magnetic charge and the external magnetic field would affect the thermodynamics. In fact, a similar
phenomena is well-known in the classical statistical physics--the external magnetic field does not
affect the classical partition function (the so-called Van Leeuwen's theorem for the non-existence
of diamagnetism in classical physics \cite{HUANG}).

It is interesting to consider the static limit of the magnetized  dipole ring solution. This limit is obtained for
$\lambda=\nu$. In the static case the conditions for absence of conical singularities are reduced to

\begin{equation}\label{SDBRCSC1}
\Delta \psi = \Delta \phi = 2\pi {(1+ \mu)^{3\over 2(1+ \alpha^2_{5}) }\over \sqrt{1-\lambda} }
\end{equation}

and

\begin{equation}\label{SDBRCSC2}
\left({1+\mu \over 1-\mu }\right)^{3\over 1+ \alpha^2_{5} }
\Lambda^{-3\over 1+ \alpha^2_{5}}|_{x=1} = \left({1 - \lambda\over 1+ \lambda} \right).
\end{equation}

The eq.(\ref{SDBRCSC2}) can be solved to determine  $B$ as a function  of the parameters $\lambda$ and $\mu$.
In other words the external magnetic field can always be chosen such that to cancel the conical singularity and
to support the static ring in equilibrium. Therefore, there exist static dipole black rings with regular horizons
in external magnetic field. This is possible because of the coupling between the external magnetic filed $B$ and
and the local charge ${\cal Q}$. The magnetized static dipole black rings provide infinite examples of regular,
static black holes with horizon topology different from that of the Schwarzschild-Tangherlini black holes,
but with same mass and asymptotics.

\section{Conclusion}

In this paper we presented explicit solutions describing magnetized black holes and  black rings in
the higher dimensional dilaton gravity. The basic physical quantities of the magnetized black objects
were calculated and some of their properties were discussed. In particular we shown that
the external magnetic field deforms the black holes horizon but it does not change
the the horizon area. Moreover, we have shown that the external magnetic field does not affect the
thermodynamics of the black objects. This seems to be related to the fact that the
electromagnetic potential is parallel to the non-rotating Killing vector which means that
there is no coupling between the rotations and the magnetic field. In the more general case of rotating Killing
vector we expect that the external field will influence the black hole thermodynamics as in the four dimensions
\cite{RADU}. The general case, however, require more sophisticated mathematical techniques and we will
address this question in a future work. We also discussed briefly the ultrarelativistic limits of the
magnetized solutions and gave an explicit example for the $D$-dimensional Schwarzschild-Tangherlini solution.
The ultrarelativistic limits of the magnetized  black holes and black rings might be useful in the
theoretical study of the black hole production in the near-future accelerators\footnote{Who knows ...?}.

\section*{Acknowledgements} The author would like to thank the  Bogoliubov Laboratory of Theoretical Physics (JINR) for their kind hospitality where part of the present work was done.
This work was partially supported by the Bulgarian National Science Fund under Grant MUF04/05 (MU-408).

\appendix

\section{Quasilocal formalism}\label{QLF}

Here we briefly discuss the quasilocal formalism in EMd gravity \cite{CGL}.
The spacetime metric can be decomposed into the form

\begin{equation}
ds^2 = - N^2dt^2 + \chi_{ij}(dx^i + N^{i}dt)(dx^j + N^{j}dt)
\end{equation}

where $N$ is the lapse function and $N^{i}$ is the shift vector.

This decomposition means that the spacetime is foliated by spacelike surfaces $\Sigma_{t}$ of metric
$\chi_{\mu\nu} = g_{\mu\nu} + u_{\mu}u_{\nu}$, labeled by a time coordinate $t$ with a unit normal vector
$u^{\mu} = - N\delta^{\mu}_{0}$. A timelike vector $\upsilon^{\mu}$, satisfying $\upsilon^{\mu}\nabla_{\mu}t=1$,
is decomposed into the lapse function and shift vector as $\upsilon^{\mu}= Nu^{\mu} + N^{\mu}$.
The spacetime boundary consists in the initial surface $\Sigma_{i}$ ($t=t_{i}$) and the final surface $\Sigma_{f}$
($t=t_{f}$) and a timelike surface ${\cal B}$ to which the vector $u^{\mu}$ is tangent. The surface ${\cal B}$
is foliated by $(D-2)$-dimensional surfaces $S^{r}_{t}$, of metric $\sigma_{\mu\nu}= \chi_{\mu\nu} - n_{\mu}n_{\nu}$,
which are intersection of $\Sigma_{t}$ and ${\cal B}$. The unit spacelike outward normal to $S^{r}_{t}$, $n_{\mu}$,
is orthogonal to $u^{\mu}$.

In order to have well-defined variational principle we must supplement
the action (\ref{EMDA}) with the corresponding boundary term :

\begin{eqnarray}\label{SEMDA}
S= {1\over 16\pi} \int d^Dx \sqrt{-g}\left(R -
2g^{\mu\nu}\partial_{\mu}\varphi \partial_{\nu}\varphi  -
e^{-2\alpha\varphi}F^{\mu\nu}F_{\mu\nu} \right) \nonumber \\ + {1\over 8\pi}\int_{\Sigma_{i}}^{\Sigma_{f}} K\sqrt{\chi}d^{D-1}x
-{1\over 8\pi}\int_{{\cal B}}\Theta \sqrt{\gamma}d^{D-1}x.
\end{eqnarray}

$K$ is the trace of the extrinsic curvature $K^{\mu\nu}$ of $\Sigma_{t_{i,f}}$ and
$\Theta$ is the trace of the extrinsic curvature $\Theta^{\mu\nu}$ of ${\cal B}$, given by

\begin{eqnarray}
K_{\mu\nu} &=& - {1\over 2N}\left( {\partial \chi_{\mu\nu}\over \partial t} - 2D_{(\mu}N_{\nu)} \right) ,\\
\Theta_{\mu\nu} &=& - \gamma^{\alpha}_{\mu} \nabla_{\alpha} n_{\nu},
\end{eqnarray}

where $\nabla_{\mu}$ and $D_{\nu}$ are the covariant derivatives with respect to the metric $g_{\mu\nu}$ and
$\chi_{ij}$, respectively.

The quasilocal energy and angular momentum are given by

\begin{eqnarray}
E = {1\over 8\pi} \int_{S^{r}_{t}} \sqrt{\sigma} \left[N(k-k_{0}) + {n_{\mu}p^{\mu\nu}N_{\nu}\over \sqrt{\chi}} \right] d^{D-2}x \nonumber \\ +
{1\over 4\pi}\int_{S^{r}_{t}} A_{0} \left({\hat \Pi}^{j} - {\hat \Pi}_{0}^{j} \right)n_{j}d^{D-2}x ,\\
J_{i} = - {1\over 8\pi} \int_{S^{r}_{t}} {n_{\mu}p^{\mu}_{i}\over \sqrt{\chi} } \sqrt{\sigma}d^{D-2}x
- {1\over 4\pi} \int_{S^{r}_{t}} A_{i}{\hat \Pi}^{j}n_{j}d^{D-2}x .
\end{eqnarray}

Here $k= - \sigma^{\mu\nu}D_{\nu}n_{\mu}$ is the trace of the extrinsic curvature of $S^{r}_{t}$ embedded in $\Sigma_{t}$. The momentum variable $p^{ij}$
conjugated to $\chi_{ij}$ is given by

\begin{equation}
p^{ij} = \sqrt{\chi}\left(\chi^{ij}K - K^{ij} \right).
\end{equation}

The quantity ${\hat \Pi}^{j}$ is defined by

\begin{equation}
{\hat \Pi}^{j} = - {\sqrt{\sigma}\over \sqrt{\chi}} \sqrt{-g}e^{2\alpha\varphi}F^{0j}.
\end{equation}

The quantities with the subscript "0" are those associated with the background. Detail discussion of the quasilocal
formalism can be found in \cite{CGL}.

Let us denote by tilde all quantities which refer to the seed solution. Then the Harrison transformation
gives new solution which is characterized with the following quantities:

\begin{eqnarray}
N &=&  \Lambda^{1\over (D-3)(1+ \alpha^2_{D})} {\tilde N}, \\
N^{\hat i} &=& {\tilde N}^{\hat i} ,\\
N^{y} &=& {\tilde N}^{y}=0 ,\\
\chi_{ij}dx^idx^j &=& \Lambda^{2\over (D-3)(1+ \alpha^2_{D})} {\tilde \chi}_{\hat{i}\hat{j}}dx^{\hat{i}}dx^{\hat{j}}
+\Lambda^{-{2\over (1+ \alpha^2_{D})}}{\tilde \chi}_{yy}dy^2, \\
\sigma_{ab}dx^adx^b &=& \Lambda^{2\over (D-3)(1+ \alpha^2_{D})}{\tilde \sigma}_{\hat{a}\hat{b}}dx^{\hat{a}}dx^{\hat{b}}
+\Lambda^{-{2\over (1+ \alpha^2_{D})}}{\tilde \sigma}_{yy}dy^2
\end{eqnarray}

where $\hat{i}$ and $\hat{a}$ take the same values as $i$ and $a$ except for $i=y$ and $a=y$. It is easy
to see that

\begin{eqnarray}
\chi &=& \Lambda^{2\over (D-3)(1+ \alpha^2_{D})} {\tilde \chi},\\
\sigma &=& {\tilde \sigma}.
\end{eqnarray}

For the cases considered in this paper we have

\begin{eqnarray}
k &=& \Lambda^{-{1\over (D-3)(1+ \alpha^2_{D})}} {\,}{\tilde k},\\
k_{0} &=& \Lambda^{-{1\over (D-3)(1+ \alpha^2_{D})}} {\,}{\tilde k}_{0},\\
K^{\mu}_{i} &=& \Lambda^{-{1\over (D-3)(1+ \alpha^2_{D})}} {\tilde K}^{\mu}_{i}.
\end{eqnarray}

Taking into account these results and the fact that in our case $A_{0}=0$ and ${\hat \Pi}^{j}=0$ we find

\begin{eqnarray}
E = {1\over 8\pi} \int_{S^{r}_{t}} \sqrt{\sigma} \left[N(k-k_{0}) + {n_{\mu}p^{\mu\nu}N_{\nu}\over \sqrt{\chi}} \right] d^{D-2}x \nonumber \\= {1\over 8\pi} \int_{S^{r}_{t}} \sqrt{{\tilde \sigma}} \left[ {\tilde N}({\tilde k}-{\tilde k}_{0})  + {{\tilde n}_{\mu}{\tilde p}^{\mu\nu}{\tilde N}_{\nu}\over \sqrt{\chi}} \right] d^{D-2}x=
{\tilde E} ,\\
J_{i} = - {1\over 8\pi} \int_{S^{r}_{t}} {n_{\mu}p^{\mu}_{i}\over \sqrt{\chi} } \sqrt{\sigma}d^{D-2}x =
- {1\over 8\pi} \int_{S^{r}_{t}} {{\tilde n}_{\mu}{\tilde p}^{\mu}_{i}\over \sqrt{{\tilde \chi}} } \sqrt{{\tilde \sigma}}d^{D-2}x = {\tilde J}_{i} .
\end{eqnarray}

Therefore the Harrison transformation leaves the quasilocal mass and angular momenta of the seed solution
unchanged. In the same way one can show that the Euclidean action (i.e. the Euclideanized version of (\ref{SEMDA}) )  of the magnetized solutions with respect to the dilaton-Melvin background $I_{P}$  coincide
with that of the corresponding seed solutions (with respect to the Minkowski background)
\begin{equation}
I_{P} = {\tilde I}_{P}.
\end{equation}

It is worth noting that the Euclideanization of the dipole black rings is subtle. In fact, as shown in \cite{AR},
we are forced to work with a complex geometry. Nevertheless, the final result gives a real action.

\section{ Ultrarelativistic limits}

The ultrarelativistic limit of the Schwarzschild-Tangherlini spacetime
is obtained via the Aichelburg-Sexl procedure, i.e. by boosting
the Schwarzschild-Tangherlini black hole to the speed of light \cite{AS}.
Performing a Lorentz boost in $z_{1}$ direction and taking the the limit $V\to 1$
while keeping the ratio $p=M/\sqrt{1-V^2}$ fixed
we obtain the ultrarelativistic limit of the Schwarzschild-Tangherlini spacetime

\begin{equation}
ds^2 = 2dud\upsilon +  dz^2_{2} + ...+ dz^2_{D-3} + d\rho^2 + \rho^2d\phi^2
+ {\cal H}\delta(u)du^2
\end{equation}

where

\begin{eqnarray}
u &=& {z_{1}-t\over \sqrt{2}} ,\,\,\, \upsilon = {z_{1}+t\over \sqrt{2}},\\
{\cal H} &=& - 8\sqrt{2}p\ln\rho \,\,\, (D=4),\\
{\cal H} &=& {16\pi\sqrt{2}p \over (D-4)\Omega_{D-3}
\left(z^2_{2} + z^{2}_{3} +...+z^2_{D-3} + \rho^2\right)^{(D-4)/2} } \,\,\, (D>4).
\end{eqnarray}

The ultrarelativistic limit of the dilaton-Schwarzschild-Melvin solution  can be found by applying the Harrison
transformation\footnote{This is possible, since, in our cases, the Aichelburg-Sexl limit
and the Harrison transformation commute as one can see.} to
the ultrarelativistic limit of the Schwarzschild-Tangherlini spacetime and the result is

\begin{eqnarray}\label{URDML}
ds^2 = \Lambda^{{2\over (D-3)(1+\alpha^2_{D})}}
\left[2dud\upsilon +  dz^2_{2} + ... + dz^2_{D-3} + d\rho^2 \right] \nonumber \\
+ \Lambda^{-{2\over 1+ \alpha^2_{D}}}\rho^2d\phi^2
+ \Lambda^{{2\over (D-3)(1+\alpha^2_{D})}}{\cal H}\delta(u)du^2 .
\end{eqnarray}

The dilaton field and the magnetic field are invariant under the Lorentz boost in
$z_{1}$-direction and remain unchanged. The metric (\ref{URDML}) represents an
impulsive gravitational wave propagating in dilaton-Melvin background (Universe) along
$z_{1}$ direction. The impulsive wave front corresponds to the null hypersurface $u=0$.
On the impulsive wave front the metric is given by

\begin{equation}
ds^2= \Lambda^{{2\over (D-3)(1+\alpha^2_{D})}}
\left[ dz^2_{2} + ... + dz^2_{D-3} + d\rho^2 \right] +
\Lambda^{-{2\over 1+ \alpha^2_{D}}}\rho^2d\phi^2 .
\end{equation}

This metric does not depend on time and, therefore, the impulsive wave is non-expanding.
In addition the explicit form of the metric shows that the impulsive wave front is curved
(for $b\ne 0$).

The ultrarelativistic limits  of the other magnetized solutions can be found in the same manner by
applying the Harrison transformation to the ultrarelativistic limits of their seed solutions
(see \cite{YOSH}, \cite{OKP1}, \cite{OKP2}).

\end{document}